# Simulating Surveys for ELT-MOSAIC: Status of the MOSAIC Science Case after Phase A

M. Puech[1], C. J. Evans[2], K. Disseau[3], J. Japelj[4], O. H. Ramírez-Agudelo[2], H. Rahmani[1], M. Trevisan[1,5], J. L. Wang[1,6], M. Rodrigues[7,1], R. Sánchez-Janssen[2], Y. Yang[1], F. Hammer[1], L. Kaper[4], S. L. Morris[8], B. Barbuy[9], J.-G. Cuby[10], G. Dalton[7,11], E. Fitzsimons[2], P. Jagourel[1], and the MOSAIC Science Team[12]

[1] GEPI, Observatoire de Paris, PSL University, CNRS, 5 Place Jules Janssen, 92190 Meudon, France
[2] UK Astronomy Technology Centre, Royal Observatory, Blackford Hill, Edinburgh, EH9 3HJ, UK
[3] Institut für Astrophysik, Georg-August-Universität, Friedrich-Hund-Platz 1, 37077 Göttingen, Germany
[4] Astronomical Institute Anton Pannekoek, Amsterdam University, Science Park 904, 1098 XH, Amsterdam, The Netherlands
[5] Instituto de Fisica, Universidad Federal do Rio Grande do Sul, Departamento de Astronomia, Campus do Vale, Av. Bent Gonçalves 9500, 91501-970 Porto Alegra, Brasil
[6] NAOC, Chinese Academy of Sciences, A20 Datun Road, 100012 Beijing, PR China
[7] Astrophysics, Department of Physics, Denys Wilkinson Building, Keble Road, Oxford, OX1 3RH, UK
[8] Department of Physics, Durham University, South Road, Durham, DH1 3LE, UK
[9] Universidade de São Paulo, IAG, Rua do Matão 1226, Cidade Universitária, São Paulo, 05508-900
[10] Aix Marseille Université, CNRS, Laboratoire d'Astrophysique de Marseille, 13388 Marseille, France
[11] RALSpace, STFC Rutherford Appleton Laboratory, HSIC, Didcot, OX11 0QX, UK
[12] See complete list at : http://www.mosaic-elt.eu/index.php/science

## ABSTRACT

We present the consolidated scientific case for multi-object spectroscopy with the MOSAIC concept on the European ELT. The cases span the full range of ELT science and require either 'high multiplex' or 'high definition' observations to best exploit the excellent sensitivity and wide field-of-view of the telescope. Following scientific prioritisation by the Science Team during the recent Phase A study of the MOSAIC concept, we highlight four key surveys designed for the instrument using detailed simulations of its scientific performance. We discuss future ways to optimise the conceptual design of MOSAIC in Phase B, and illustrate its competitiveness and unique capabilities by comparison with other facilities that will be available in the 2020s.

## 1. INTRODUCTION

The 39m Extremely Large Telescope (ELT) will be the largest optical-near infrared telescope in the coming decades. It will be equipped with a suite of instruments which span the parameter space required to address a large collection of science cases that were developed over the past 15 years[1]. In particular, the detailed science case for a multi-object spectrograph (MOS) on the ELT was initially developed in the context of the ESO Design Reference Mission (e.g., Puech et al. 2008; 2010) and during Phase A studies of the EAGLE (e.g., Cuby et al. 2010, Evans et al. 2010) and OPTIMOS (Navarro et al. 2010; Le Fèvre et al. 2010) concepts. More recently, the case for an ELT-MOS was assembled from consultation with the European user community (and beyond), as presented in the ELT-MOS White Paper (Evans et al. 2013), with a significant update by Evans et al. (2015).

The White Paper formed the initial core of the MOSAIC science case, and was used to inform the top-level requirements (TLRs) for the conceptual Phase A study, undertaken between March 2016 and March 2018. Detailed scientific simulations with the WEBSIM-COMPASS software (Puech et al. 2016) were used to assist with science trade-offs in the study, while several new science cases were also identified and added to the initial list. The Science Team prioritised four highlight cases for future MOSAIC observations, and defined potential surveys that were simulated and dimensioned using the MOSAIC conceptual design, which was finalised in parallel by the technical team.

In this paper we summarise the MOSAIC science case at the end of Phase A and highlight some of the science trade-offs in the design. We then introduce potential Public Surveys for the first years of operations and the simulations conducted to define them. The overall status of the MOSAIC project is given in these proceedings by Morris et al., with the conceptual design presented by Jagourel et al. and the science specification and operational concept presented by Sánchez-Janssen et al.

---

[1] https://www.eso.org/sci/facilities/eelt/science/doc/eelt_sciencecase.pdf

# 2. STATUS OF THE MOSAIC SCIENCE CASE

## 2.1. Science trades during Phase A

Building on the cases assembled from the White Paper, the first science activities in Phase A focussed on the flow-down of instrument requirements from each case and then consolidation into the TLRs for the technical design (Sánchez-Janssen et al. these proceedings). In defining the instrument TLRs, it was clear from the outset that it would not be possible to accommodate all of the capabilities desired by the Science Team within a feasible design concept and affordable hardware budget. Three detailed aspects of the TLRs were examined early in the study, with the conclusions now briefly summarised (for further details see Evans et al. 2016).

- **Inclusion of *K*-band**: this would open-up exciting and novel observations but is unrealistic within the current financial constraints. This rendered three original cases from the White Paper as infeasible: the study of early-type galaxy clusters, and study of the Galactic central parsec and its central star cluster. Several other cases become only partially feasible, but observations at shorter wavelengths will still satisfy some of the scientific motivations. For example, the lack of *K*-band limits the reach of extragalactic science, ruling out studies of galaxy metallicities and dynamics at $z \geq 4$ and the quest for observations of the first galaxies at $z \sim 10\text{-}12$ using HeII and CIII]. Studies of the Ly$\alpha$ escape fraction is feasible using H$\beta$ alone, so coverage of the *H*-band still extends this out toward $z \sim 3$.

- **Blue-visible performance**: The combined efficiency of the ELT and MOSAIC at <0.4µm is sufficiently low that a blue-optimised instrument on the VLT could potentially be competitive (principally due to the properties of the proposed coatings of the ELT mirrors). Given the large parameter space demanded for MOSAIC observations, we therefore limited the blueward coverage of the visible spectrographs to $\geq 0.40$µm (Evans et al. 2016), updated to $\geq 0.45$µm (from similar efficiency comparisons) as the study progressed; the exact blueward cut-off will be revisited in Phase B.

- **Motivations for high spectral resolution in the near IR**: A high-resolution (HR, $R$>18,000) mode in the *H*-band was adopted as an interesting addition to the initial TLRs. This is primarily motivated by studies of stellar populations – building on work with the APOGEE and MOONS surveys – but also with applications in studies of high-redshift galaxies. Analysis of the spectral resolution in the near-IR required for stellar abundances was performed by Dr. O. Gonzalez (priv. comm.) in the frame of science simulations for the HARMONI instrument (see Clarke et al. these proceedings) but which are also relevant for MOSAIC. The conclusion was that the minimum resolution for estimates of [O/Fe] is $R \sim 17,000$, but higher resolution is needed to reduce uncertainties toward the desired ±0.1 dex.

## 2.2. MOSAIC Science Case before prioritisation

Taking into account the above decisions re: wavelength coverage, the MOSAIC science case spans six broad topics, which have been revised and extended during Phase A compared to the original cases described in the White Papers.

- **SC1. 'First light' – spectroscopy of the most distant galaxies:** Exactly when and how reionisation occurred is still unknown. Identification of the main (and elusive) ionizing sources requires detailed observations of the Ly$\alpha$ properties of faint objects. This will enable a precise characterisation of the ionisation state of the intergalactic medium (IGM) in the first Gyr ($5 < z < 13$), allowing us to reconstruct the timeline of reionisation. MOSAIC will provide the largest observational sample of first galaxies at sufficient spectral resolving power to determine the properties of their stellar populations and their interstellar media (ISM), and to search for the presence of outflows.

- **SC2. Evolution of large-scale structures:** Studies of large-scale structures provide the means to link galaxies and gas with the underlying density field. The high-redshift IGM contains most of the baryons in the Universe, and is therefore the mass reservoir that feeds galaxy growth. The complex interplay between galaxies and gas, occurring on Mpc scales, is central to models of galaxy formation. The goal is to reconstruct the 3D density field of the IGM at $z \sim 3$ on these scales to study its topology, its chemical properties, and to correlate the position of the galaxies with the density peaks. Furthermore, the inventory of baryons shows that they appear to be missing in galaxy halos when compared to the universal fraction (Fukugita et al. 1998; McGaugh et al. 2010). Recent observations of the circumgalactic medium (CGM) in galaxies at $z\sim0.2$ may have identified these 'missing baryons' as they show the added contribution of cold/warm/hot diffuse gas in the halo could be up to six times that of the stars and neutral gas (Werk et al. 2014), albeit with large uncertainties. MOSAIC will probe the different gas phases (cold vs. warm vs. hot) of the CGM in individual distant halos at $z\sim3$.

- **SC3. Mass assembly of galaxies through cosmic time:** Advancing the field of galaxy formation requires a comprehensive census of the mass assembly, star-formation histories, and stellar populations in galaxies. MOSAIC will deliver several advances in the field by measuring the stellar kinematics in $z\sim1$ galaxies or in sub-structures of local systems. In particular, measuring the stellar kinematics of the tidal debris expected to lie in the outskirts of most Milky Way-like systems will allow us to test theoretical predictions and learn about the processes by which these galaxies formed (Toloba et al. 2016), and provide unique views on their dark matter halos (Errani et al. 2015). At earlier epochs ($2<z<4$), it will be important to measure spatially-resolved chemo-dynamical information in galaxies across a wide range of stellar masses (dwarfs to giants) and environments (field to clusters). Such observations will allow us to accurately measure the evolution of the fraction of rotationally-supported galaxies vs. mergers out to $z\sim4$, to study the evolution of important scaling relations and quantities such as the Tully–Fisher relationship or velocity dispersion, and to derive the internal distribution of metals. Going one step further, while rotation curves at the outskirts of spirals have been extensively studied in the local Universe, first attempts at $z\sim1$-2 have just been obtained at the VLT (Genzel et al. 2017), and measurements at $z>2$ are completely new territory. Measuring accurate rotation curves will enable us to infer the dark matter content in statistically meaningful samples of galaxies out to $z\sim4$, providing important tests of structure formation within the $\Lambda$CDM paradigm, and in particular of cosmological simulations.

- **SC4. AGN/galaxy coevolution and AGN feedback:** Understanding the combined evolution and growth of galaxies and their central super-massive black holes (SMBHs) is crucial to the field of galaxy formation. A key question is how galaxy and SMBH growth is self-regulated by the feedback processes from active galactic nuclei (AGN) and supernova-driven outflows, which can heat up the host galaxy ISM and expel it from the halo. Such outflows are thought to be behind the termination of massive galaxy growth but are still poorly understood. As well as characterising outflows at the peak of AGN activity, we need to constrain the AGN occupation distribution. Large spectroscopic samples of AGN will shed light on scenarios of BH seed formation, and quantify the role of AGN in heating the high-$z$ IGM, assessing their contribution to reionisation.

- **SC5. Resolved stellar populations beyond the Local Group:** Resolved stellar populations enable us to explore the star-formation and chemical-enrichment histories of galaxies, providing direct constraints on galaxy formation and evolution models. Studies in the Local Group have shown that precise chemical abundances and stellar kinematics can break the age-metallicity degeneracy, while also helping to disentangle the populations associated with different structures. The ELT will bring a wealth of new and exciting target galaxies within our grasp for the first time – spanning a much broader range of galaxy morphologies, star-formation histories and metallicities than those available to us at present – enabling us to test theoretical models of galaxy assembly and evolution in systems out to distances of several Mpc.

- **SC6. Galaxy archaeology:** A key case in this area is to investigate the existence of the `Spite Plateau' (Spite & Spite, 1982) in the Li abundances of metal-poor stars in extragalactic systems. The most straightforward interpretation of the plateau in the Galaxy is that the observed Li was primordial. In those early moments, only nuclei of deuterium, two stable helium isotopes ($^3$He, $^4$He) and $^7$Li were synthesized, with their abundances dependent on the baryon/photon ratio, i.e. the baryonic density of the Universe. Thus, the Spite Plateau provides an estimate of the baryon/photon ratio. Measurements of the baryonic density (with unprecedented precision) from fluctuations of the cosmic-microwave background by *WMAP* (Spergel et al. 2007) have challenged this interpretation, but there are indications that the plateau is universal, with Monaco et al. (2010) finding evidence for it in ω Cen (generally considered the nucleus of a disrupted satellite galaxy). It is therefore important to test whether this is the case in other Local Group galaxies, necessitating high-resolution spectroscopy ($R>20,000$) in a system such as the Sagittarius Dwarf ($I\sim21.5$ mag).

All of the above cases developed for MOSAIC can be done using the Phase A design except for SC6. The decision early in the study to limit the blueward coverage (Sect. 2.1) had an immediate impact on some of the cases previously advanced for SC6 in the White Papers as many of the diagnostics commonly used to study metal-poor stars are now inaccessible (e.g. Ca *K*, CH *G*-band, Sr II lines). In contrast, the Li case summarised above requires high-resolution spectroscopy of the Li I 6708Å doublet, which remains available. Unfortunately, the performance of the high-resolution setting in the visible spectrograph at the end of Phase A was only partially compliant in terms of wavelength coverage (with >30% dispersion efficiency) and moderately lower resolution ($R$~15,000-17,000); this is primarily limited by the size of the optics, so further reoptimisation is not expected to yield substantial gains. In short, with the current design, the Li case appears challenging and requires quantitative simulations ahead of Phase B to better assess its feasibility.

### 2.3. MOSAIC observing modes

To satisfy the assembled science cases and requirements, the Phase A design includes four observational modes, each with sufficient multiplex to be competitive with other planned facilities:

- **High-definition mode (HDM):** Simultaneous observations of 8 IFUs (goal: 10 IFUs) deployed within a ~40 arcmin$^2$ patrol field and each with enhanced image quality from multi-object adaptive optics (MOAO, see T. Morris et al. in conf. 10703). Each IFU will cover a 1.9 arcsec hexagon with 80 mas spaxels, with the spectrographs delivering $R$~5000 over 0.8-1.8μm (between 250 and 430nm in one observation). A high spectral-resolution set-up will provide $R$~20,000 over a passband of ~100nm at ~1.6μm. The light entering each spaxel will be injected into a fibre that will be imaged onto 3.06 pixels on the detector.

- **Visible Integral Field Unit mode (VIFU):** Simultaneous observations of 8 IFUs (goal: 10 IFUs) deployed within a ~40 arcmin$^2$ patrol field with telescope delivered 'seeing-limited' performance (or GLAO-corrected, using M4 of the telescope). Each IFU will cover a 2.3 arcsec hexagon with 138 mas spaxels, with the spectrographs delivering $R$~5000 over 0.45-0.92μm (between 150 and 230nm in one observation). A possible high spectral-resolution set-up with R~15,000 could allow coverage of a passband of ~60nm around 0.65μm and/or 0.86μm. The light entering each spaxel will be injected into a fibre that will be imaged onto 4.21 pixels on the detector.

- **High multiplex mode in the near-IR (HMM-NIR):** Simultaneous integrated-light observations of 80 objects (goal: 100) with dedicated sky fibres operated in cross-beam switching sequences (see Sánchez-Janssen et al. these proceedings; see also Rodrigues et al. 2010; Yang et al. 2012; Puech et al. 2012). Each object will be observed with a bundle of 19 x 100 mas fibres, giving an on-sky aperture of 500 mas in diameter. Each fibre will be imaged onto 3.06 pixels on the detector. The spectrograph setups are identical to the HDM mode.

- **High multiplex mode in the visible (HMM-Vis):** Simultaneous integrated-light observations of 80 objects (goal: 100). Note that there will be no dedicated sky fibres in this mode, as the sky background is much fainter than in the near-IR. Each object will be observed with a bundle of 19 x 168 mas fibres, giving an on-sky aperture of 840 mas in diameter. Each fibre will be imaged onto 4.21 pixels on the detector. The spectrograph setups are identical to the VIFU mode.

### 2.4. Prioritising science cases and identifying MOSAIC surveys

Several science meetings were held during the Phase A study to discuss the relevant trades and prioritisation, culminating in discussions on which cases the community sees as most transformational at a conference in Toledo in Oct. 2017 (Evans et al. 2018). Four highlight cases were identified that will help drive future decisions on instrument capability and will form the basis for future surveys (either as a large part of the Guaranteed Time Observations (GTO), separate Large Programmes, and/or Public Surveys). These are summarised in Table 1, which includes the primary observational mode(s) for each and those that will be beneficial but not essential.

Table 1: High-priority MOSAIC surveys (not in a ranked order).

| Survey & related Science Case(s) | Primary/essential (secondary/beneficial) modes |
|---|---|
| 1. Evolution of dwarf galaxies<br>    SC2: mass assembly; SC1: contribution to reionisation | HDM (HMM-NIR) |
| 2. Inventory of matter<br>    SC2: IGM tomography; SC2: missing baryons; SC3: dark matter profiles in high-z galaxies | HDM + HMM-Vis (VIFU) |
| 3. First-light galaxies<br>    SC1: the sources of reionisation | HMM-NIR (HDM) |
| 4. Extragalactic stellar populations<br>    SC5: evolved populations beyond the Local Group | HDM (HMM-NIR) |

## 3. SIMULATING MOSAIC SURVEYS

### 3.1. Science simulations and metrics

We used the WEBSIM-COMPASS scientific simulator, which builds on the previous series of WEBSIM simulators developed during the ESO E-ELT Design Reference Mission and Instrument Phase A studies. The WEBSIM-COMPASS simulator consists of a web interface coupled to an IDL code, which allows the user to perform end-to-end simulations of ELT observations, and in particular with MOSAIC. The simulation pipeline produces fake observations in FITS format that mimic the result of a data-reduction pipeline with perfectly extracted/reduced data. See Puech et al. (2016) and references therein for further details.

WEBSIM-COMPASS was used to simulate future MOSAIC programmes and to estimate the total observing time required to complete different surveys. The simulations were first used to check in which regime these programmes will be conducted and establish scaling relations between signal-to-noise ratio ($S/N$) and different parameters such as IFU spaxel scale, spectral resolution ($R$), or *Throughput*. We checked that in most foreseen observations these scaling relations are those expected from a (background) shot-noise limited regime:

$$S/N \propto \frac{IFU\ Spaxel\ Size \times Telescope\ Diameter \times \sqrt{Throughput} \times \sqrt{Observing\ Time}}{\sqrt{R}}$$

To allow a meaningful comparison between programmes, observing modes, and/or facilities, our adopted metric was survey speed, which can be defined as the ratio between the instrument *Multiplex*, which is the number of on-sky apertures that can be used to conduct simultaneous observations, and the total observing time required (to achieve the $S/N$ necessary for measurements on all targets). Thus, the larger the Survey Speed, the more efficient/faster the instrument. Given the above scaling relation between $S/N$ and *Observing Time*, one gets the following useful scaling relations:

$$Survey\ Speed = \frac{Multiplex}{Observing\ Time} \propto D^2 \times Throughput \times Multiplex$$

Alternative metrics to compare facilities have been proposed, such as the *Etendue* or the so-called $A\Omega$ product (the product between the telescope collecting surface $A$ and the solid angle $\Omega$ of the instrument field-of-view). However, this is not well-suited for comparing facilities on telescopes with drastically different sizes as the *Etendue* does not account for sensitivity. Accounting for the *Etendue* (or proxies) alone can therefore lead to misleading conclusions (see discussion in Sect. 4.5). The time required to achieve a given $S/N$ is a better metric for comparing a given science case on different facilities, including varying apertures, as it accounts for all factors. In this context, *Survey Speed* as defined above can be interpreted simply as the inverse of the average time per target (accounting for differences of *Multiplex*) required to achieve the necessary $S/N$.

## 3.2. Survey 1: Mass assembly and evolution of dwarf galaxies

Sub-M* (galaxies with masses smaller than the knee of the luminosity/mass function) and, in particular, dwarf galaxies are expected to play a key role in galaxy formation and evolution. In hierarchical models they are thought to be the first structures to form in the Universe and are believed to have an important contribution to the reionisation process. Investigating the detailed properties of dwarf galaxies and their relation to the more massive population at z>2 is therefore an important test of structure formation in ΛCDM. Distant sub-M* galaxies have faint apparent magnitudes and we have limited knowledge of their morphological and spectroscopic properties (e.g., Kassin et al. 2012; van der Wel et al. 2014; Kartaltepe et al. 2015; Whitaker et al. 2015; Simons et al. 2015). The samples for which spatially-resolved kinematics can be obtained remain small and the integration times very large. For instance, Contini et al. (2016) assembled a sample of 28 galaxies at 0.2<z<1.4 with stellar masses in the range ~$10^8$-$10^{10}$ $M_\odot$ using VLT-MUSE, but this required a total exposure of 27 hr. This domain will probably remain only partially explored until the next generation of integral-field spectrographs on the ELTs (and *JWST*).

Connecting the distant dwarf population to the more massive galaxies will require mass-representative observational samples covering a large range in mass. Focusing on redshift bins at *z*=2 (for continuity with on-going IFU surveys with VLT-KMOS) and *z*=4, we can sample each epoch using at least three bins in stellar mass (sampling sub-M*, M*, and super-M* galaxies at both redshifts, see Puech et al. 2010). Furthermore, at least three different fields will have to be sampled at each epoch/mass to average fluctuations associated with cosmic variance. The survey will target emission lines from galaxies such as Hα or [OII] between the OH sky lines in either the *J* or *H* bands. The survey will therefore require pre-existing deep imaging down to $H_{AB}$~25 with good spectroscopic completeness, within reach of facilities such as *JWST*-NIRCam or *Euclid* (imaging) and VLT-MOONS or *JWST*-NIRSpec (spectroscopic redshifts).

The first objective of the MOSAIC survey will be to measure the spatially-resolved kinematics of *z*=2-4 galaxies using the HDM IFUs (see Fig. 1). When combined with deep imaging (rest-frame near-IR) which traces galaxy morphology, we will infer the dynamical state of distant galaxies and investigate the evolution of the fraction of virialised rotating disks vs. unvirialised systems such as mergers (see, e.g., Rodrigues et al. 2017 and references therein). Using kinematics or morphology alone can result in large systematics in the inferred fraction of virialised rotating disks. To have such a survey with the internal statistics of each bin limited only by Poisson fluctuations (and not by systematics associated with analysis methods), we need at least 50 galaxies per bin, which will deliver a precision of ~14% in each bin (~$1/\sqrt{50}$), commensurate with the typical accuracies of the morphological and kinematic classifications. The resulting total number of galaxies in the survey is ~1000, i.e., 50 x 3 (# of fields) x 2 (# of redshift bins) x 3 (# of mass bins).

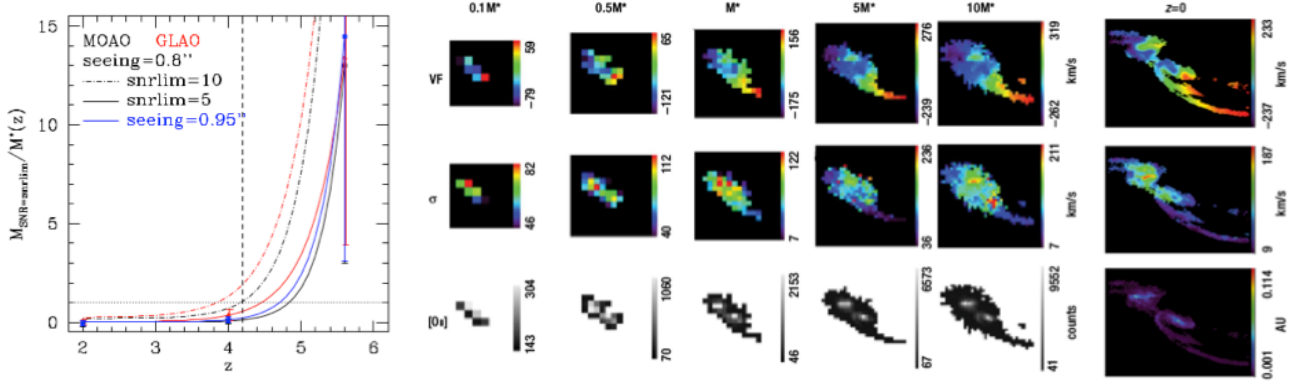

**Fig. 1**: *Left*: minimal stellar mass (in fraction of M* at a given *z*) that can be reached for different limits on the spatially-averaged *S/N* obtained in the observed emission line (*snrlim*) as a function of redshift, AO correction, and seeing (from Puech et al. 2010). For a spaxel scale of 80 mas, these limits can be reach with typical integrations of ~10 hr on source. Error-bars represent simulations for a range of morpho-kinematic templates sampling different galaxy types (from on-going major mergers to rotating disks). The vertical dashed line shows the limit above which the [OII] emission line is redshifted into the *K* band (into the more challenging regime for the thermal background and not covered by MOSAIC). A minimal average of *S/N* ~5 is required to study the spatially-resolved kinematics in distant galaxies (see Puech et al. 2010). *Right:* simulated velocity fields of a *z*~4 major merger (Puech et al. 2010) with the input high-resolution template on the right-hand side. The simulations are shown as a function of the characteristic galaxy mass (M*), with the upper panels showing the recovered velocity field; central panels: velocity dispersion; lower panels: [OII] emission-line map. Both panels show that spatially-resolved kinematics can be measured in sub-M* galaxies out to *z*~4.

The dynamical state of galaxies is mostly imprinted on large-scale motions and depends weakly on the small-scale irregularities that are due to non-circular motions such as bars or warps (Puech et al. 2008). With a spaxel scale of ~80 mas, the MOSAIC/HDM IFUs will provide at least two spaxels per half-light radius in even 0.5M* galaxies at $z$=4 (Puech et al. 2010), which allows sampling of the large-scale motions over a large range of mass between $z$=2 to 4. The *S/N* required to obtained IFU observations of galaxies as a function of mass and redshift was studied in detail during the E-ELT DRM (Puech et al. 2010). Adapting those results to the above survey, one finds that MOSAIC will be able to conduct this survey in ~125 nights (assuming 8 hr of observations per night and accounting for 30% overheads). This scales linearly with multiplex so that, for instance, decreasing/increasing the multiplex in HDM down/up to 4/10 would require 250/100 nights. This suggests that at least 8 IFUs in HDM are required to remain effective, and that increasing the number of IFUs from 8 to 10 would be highly desirable as it provides a direct gain of 20% in survey speed.

The spatially-resolved spectroscopy of sub-M* galaxies at $z$=4 at sufficient spectral resolving power to resolve their internal motions will be a unique case for MOSAIC cf. *JWST*-NIRSpec. The latter will provide measurements at similar spatial scales (~100 mas) but will be limited to $R$~2700, i.e., a velocity resolution of 110 km/s, cf. 60 km/s with MOSAIC. The HR spectral set-up ($R$=20,000) will allow us to resolve emission lines down to ~15 km/s in specific redshift windows, enabling motions to be resolved in distant LMC-like dwarfs at $z$~1.3-1.5 and $z$~3.1-3.4 using Hα or [OII], respectively. Preliminary simulations show that mapping the properties of dwarf starbursts similar to Haro11 at $z$~2 should be feasible in ~10hr with HDM observations. Further simulations are required to assess the possible *S/N* for such HR observations and to what extend spatially-resolved motions with HDM observations could be obtained in such faint and distant objects vs. integrated-light measurements with the HMM-NIR mode. For the latter mode, the number of sub-M* galaxies with $M_{stellar}$<3x10$^9$M$_\odot$ between $z$=0.8 and 2.4 is estimated to be ~800 (1750) down to $J_{AB}$=26 (27) per MOSAIC field. Such observations will provide important constraints on the star-formation history (e.g., Pacifici et al. 2012) of this population, as well as estimates of star-formation rates and metallicities from their emission lines.

This case provides strong synergies with HARMONI as several sub-samples drawn from the MOSAIC parent survey could be followed-up at higher spatial resolution to study, e.g., the non-circular motions occurring at smaller spatial scales (e.g., bars, warps) or instabilities such as clumps, which might play an important role in galaxy evolution and formation. For instance, simulations by Zieleniewski et al. (2015) have shown that such irregularities within the optical radius of $z$~2 galaxies can be recovered with HARMONI using 20 mas spaxels and integrations of 10 hr.

### 3.3. Survey 2: Inventory of matter

One of the most ambitious surveys foreseen for MOSAIC will combine both visible and near-IR observations to conduct the first direct inventory of matter in distant galaxies at $z$~3, including characterising the dark matter profiles in disk galaxies, the distribution of neutral gas in the IGM, and probing all gas phases in the CGM, as detailed below.

### 3.3.1. Dark matter profiles in distant galaxies

Dark matter profiles can be estimated from accurate measurements of rotation curves (RC) in disk galaxies. This requires sufficient spatial resolution to resolve the shape of the rotation curve (particularly the inner part) and to avoid strong biases due to beam-smearing effects, as well as sufficient *S/N* out to at least two optical radii so that the plateau of the RC can be measured accurately to within a few percent (Bosma 1978; Epinat et al. 2010). Rotation curves can be measured on individual disk galaxies but binning can be used to increase *S/N* and smooth out the small-scale fluctuations associated with non-circular motions that are unrelated to the underlying mass distribution (e.g., bars, warps, etc.). First attempts to measure rotation curves in distant galaxies using binning were just obtained in galaxies at $z$~2 at the VLT (e.g., Genzel et al. 2017), although with limited spatial extension and spatial resolution. In the local Universe, binning has been used to sample the luminosity function over ~7 magnitudes (Persic et al. 1996; Salucci et al. 2007; Lapi et al. 2018). The goal of this MOSAIC survey is to obtain similar information in galaxies at $z$=2 to 4, for the first study of the evolution of dark matter content as a function of mass and redshift. This will provide new and important tests of structure formation in the ΛCDM paradigm and in cosmological simulations, such as the evolution of the stellar vs. halo mass relationship, the evolution of the star-formation efficiency, and the evolution of disk and halo angular momentum and whether these are conserved during disk formation as predicted (see, e.g., Lapi et al. 2018).

This survey will require a representative parent sample of galaxies which sample the galaxy mass function between $z$=2 and 4. To be meaningful, the RC measurements and dark-matter profile analysis have to be conducted in the sub-sample of galaxies that are truly rotating and virialised so that the observed kinematics can be safely related to the underlying mass distribution (which is not necessarily the case in, e.g., on-going mergers). The mass-assembly survey (Sect. 3.2) will provide a natural parent sample for this programme, from which a representative sub-sample of secure distant virialised disk galaxies (VDGs) can be extracted. The fraction of such VDGs was estimated to be ~50% at $z$~0.6 (Hammer et al. 2009) and ~25% at $z$~0.9 (Rodrigues et al. 2017), but remains poorly constrained at $z$>1. An upper limit is given by the fraction of galaxies that have their kinematics dominated by a single velocity gradient (compared to velocity dispersion), which was found to be decreasing from ~ 80% at $z$~0.9 to ~35% at $z$~3.4 (see Turner et al. 2017). Adopting a hypothetical, conservative fraction of ~25% virialised rotating disks at $z$>2, the parent survey would offer at least ~250 VDGs over $z$=2 to 4 in which the decomposition of the RC into mass profiles could be conducted.

Some of the targets might require follow-up beyond the observations in Survey 1 to reach sufficient *S/N* in their outer regions. In this case, simulations show that the 80 mas spaxel scale of MOSAIC will provide the required *S/N* in 2 hr per target: an additional ~10 nights would be sufficient to provide the data needed to analyze ~100 VDGs at $z$~4 (see Fig. 2). These observations will also require good extended photometry (with *JWST*-NIRCam or ELT-MICADO) to measure the optical radii of distant galaxies with 10% accuracy and to identify those with morphological types later than Sb to minimize the impact of the bulge on the rotation curve decomposition into mass profiles (see Lapi et al. 2018).

To reach a similar level of precision/accuracy compared to studies in the local Universe, ~0.1 dex accuracy on the average rotation curves obtained in each redshift/mass bin will be required (Lapi et al. 2018). Assuming that the uncertainties in each bin are dominated by Poisson-noise fluctuations due to the limited number of galaxies per bin (and not by any error associated to, e.g., kinematic measurements, beam-smearing effects during the rotation curve modeling, or sample selection, see above) at least ~50 galaxies per bin will be required, which translates into a precision of ~0.15 dex (see Sect. 3.2). To construct an average RC in three bins of mass at two redshifts will therefore require a total sample of ~ 300 VDGs at $z$=2-4, consistent with the expected (conservative) fraction of distant VDGs estimated above.

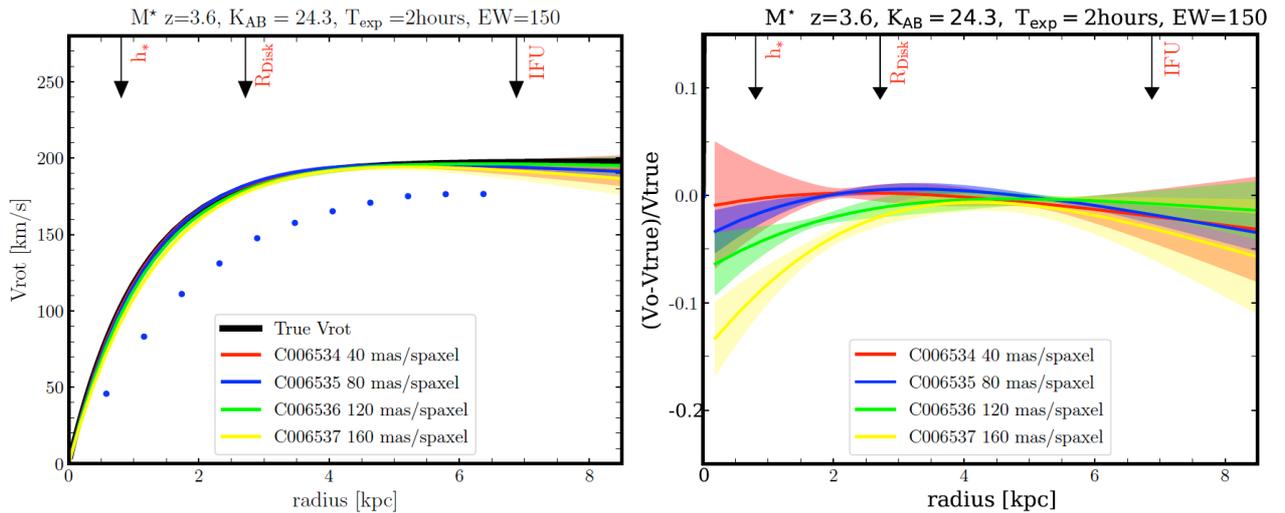

**Fig. 2:** Example of a rotation curve measured in simulated HDM observations of a $z$=3.6 distant virialised rotating disk. *Left:* rotation curve estimated using Galpak3D (Bouché et al. 2015) for different spaxel scales. The input rotation curve template is shown in black, while the rotation curve extracted from a pseudo-slit aligned to the galaxy major axis is shown as blue dots. The latter curve strongly underestimates the true rotation due to beam-smearing effects. Galpak3D is able to correct for this effect provided that the PSF can be estimated and that observations provide enough *S/N* and spatial resolution. *Right:* residual of the fitted rotation curve compared to the input template (black curve in left-hand panel). Reliable estimates for the rotation curve require spaxel sizes no larger than ~ 80-120 mas to provide at least *S/N* > 3 out to twice the optical radius, sampled by at least ~3 spatial-resolution elements (~6 spaxels); the spaxel size of 80 mas of MOSAIC satisfies these requirements for galaxies at $z$~4 in ~2 hr integrations.

### 3.3.2. Tomography of the IGM

This programme will observe the Ly$\alpha$ forest imprinted on the spectra of Lyman-break galaxies (LBGs) at $z$~3-4 over a large enough area to overcome cosmic variance of the IGM (~1 deg$^2$). The absorption lines of the Ly$\alpha$ forest allow us to trace the neutral HI gas in the IGM along the lines of sight provided by LBGs (Fig.3, left panel). Using a tomographic analysis, observing all the lines of sight offered by LBGs over 1 deg$^2$ will allow us to reconstruct the 3D distribution of the neutral HI gas in the IGM at scales of ~3-5 Mpc. To date, this technique has been applied using bright QSOs as background sources, resulting in larger reconstructed scales, or using LBGs at $z$~2.3-2.6 as background sources, but from low *S/N* spectra and over a small area on the sky (Lee et al. 2014, 2017).

This programme will require pre-identification of LBGs at $z$~3 from multi-band imaging from existing photometry or, e.g., *JWST*-NIRCam. We estimated that within one MOSAIC field of view (~40 arcmin$^2$), there should be, on average, 30 to 40 LBGs between $z$~3 and 3.5 down to $R_{AB}$=25.5 mag. Simulations show that the MOSAIC VIFUs can provide enough *S/N* to detect the Lyα forest in ~10 hr of observations once (re-binned to *R*=2000, or in 20 hr at *R*=5000), as shown in Fig. 3 (right panel). Covering 1 deg$^2$ will require ~90 MOSAIC pointings so that with 10 VIFUs, ~110n will be required to complete this programme. However, this would result in an incomplete survey as only part of the population of LBGs at z~3 can be observed, limiting the tomographic reconstruction to larger spatial (transverse) scales. A complete survey (effectively sampling scales of ~3-5 Mpc) could be obtained by using the HMM-Vis fibre bundles. In this case, the *S/N* obtained is reduced by 20% compared to the VIFU at constant integration time, so that ~15 hr of integration per target would be required to reach the same *S/N*. The resulting total observing time for this programme would then be ~170n. The latter is directly driven by the size of the reconstructed transverse scale, which in turn requires the detection of all LBGs at z~3 within each MOSAIC field. If the size of the reconstructed scale is relaxed, then we can target only a sub-sample of brighter LBGs. For instance, down to $R_{AB}$=25 mag (0.5 mag brighter), there should be, on average, ~15 LBGs per MOSAIC field, which would need integration times around a factor of three smaller to reach similar *S/N* (see Fig. 3). The resulting survey time would then reduce from 170 down to ~ 57n.

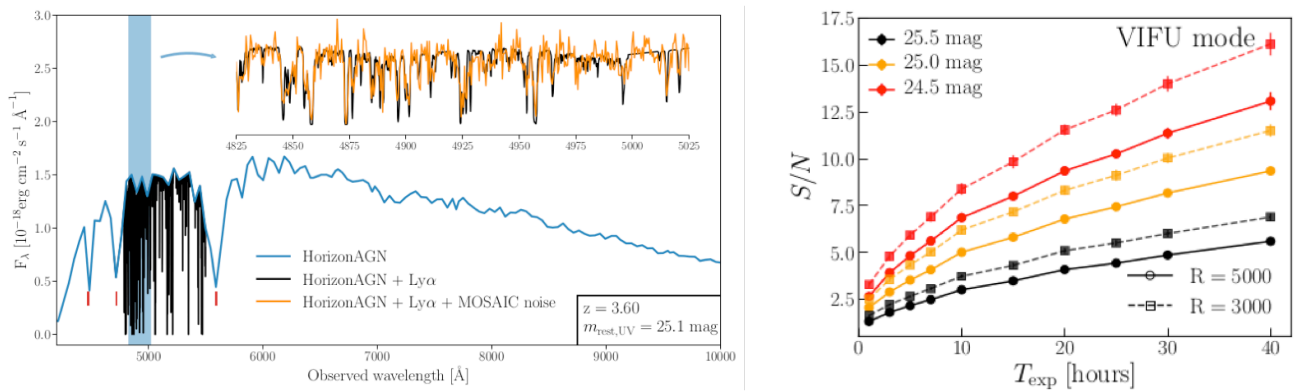

**Fig. 3:** *Left:* Example of a galaxy spectrum simulated in the Horizon-AGN simulation (Laigle et al., in prep.) and used as a template for simulating MOSAIC observations. The galaxy lies at $z$~3.60 and has a magnitude of $m_{rest,UV}$ = 25.1 mag; its synthetic spectrum is shown in blue. Red dashes indicate the positions of Lyα, Lyβ and Lyγ lines originating in the galaxy. The spectrum with the added Lyα forest (only in the region between Lyβ and Lyα lines of the galaxy) is shown in black. Realistic Gaussian noise is added to this spectrum and the resulting spectrum (here shown for the fiducial case of *R*=5000 and exposure time of 15 h) is shown in orange in the inset plot. *Right: S/N* as a function of exposure time for galaxies of three different brightnesses at $z$~3.3. *S/N* is estimated from integrated spectra obtained from VIFU simulated observations at two spectral resolving powers (see Japelj et al., in prep.).

Tomography of the IGM will also be an important goal of the other 20-30m class telescopes. For instance, WFOS on the TMT will offer a multiplex of several hundred in the visible (Bundy et al., these proceedings). Next-generation MOS instruments on 4-8m class telescopes have much larger multiplexes that can reach thousands of objects, so a high-multiplex MOS on the VLT dedicated to surveys could perhaps be a viable alternative. In Table 2 we derive the survey speed ratio between ELT-MOSAIC and these two other facilities. In this comparison, we accounted for the fact that the effective multiplex provided by these instruments for this science case will be limited by the cumulative source density of LBGs at $z$=3-3.5 down to $R_{AB}$=25.5, which is ~ 0.9 source/arcmin$^2$.

Table 2 suggests that a highly-multiplexed MOS on the VLT would be ~1.7 times faster than MOSAIC at conducting this programme, while a visible MOS on the TMT would be somewhat faster. In this comparison, we neglected variations in on-sky aperture sizes and spectral resolutions, and we assumed that the required *S/N* can be reached by all facilities: since ~10 hr of observing time per target will be required with MOSAIC to reach a *S/N*~3 per element of spectral resolution at $R=2000$, the same $S/N$ would require ~100 hr (150 hr for $R=5000$) per target on an 8m telescope under the same assumptions. Integrating over such a large number of exposures requires special care of potential sources of systematics (e.g., scattered light in the instrument, data reduction process, etc.) and it has not been demonstrated yet that the required accuracy on such extended spectroscopic observations could be reached. A MOS on the TMT with good throughput would, in principle, have an advantage in survey speed since the MOSAIC throughput is limited to ~15% because of the relatively low global transmission of the ELT at these wavelengths[2]. Given the rough comparison conducted in Table 2 we conclude that ELT-MOSAIC and a MOS on TMT should be more or less comparable in terms of survey efficiency for this science case, while a high-multiplex (visible) MOS on a dedicated 8m telescope could benefit from an advantage if very long exposures (>100 hr) can be conducted. As discussed above, these large integration times are driven by the size of the reconstructed transverse scale: in practice, reconstructions at scales of ~3-5 Mpc will probably remain limited to instruments such as MOSAIC on 20-40m telescopes.

**Table 2:** Survey speed comparison between MOSAIC HMM-VIS and other Visible MOS.

|  | D [m] | Typical visible throughput [%] | Effective Multiplex for LBGs (and patrol fields) | Approx. Survey Speed MOSAIC / other | Required observing time per target [hr] |
|---|---|---|---|---|---|
| **MOSAIC** | 39 | 15* | 36 (40 arcmin$^2$) | 1.0 | 10 |
| **VLT** | 8 | 35 | 630 (700 arcmin$^2$) | 0.6 ($\pm$0.2) | 100 |
| **TMT** | 30 | 35 | 37 (40.5 arcmin$^2$) | 0.7 ($\pm$0.2) | 7 |

*ultimately limited by transmission of the telescope.

### 3.3.3. Missing baryons in the CGM

Identifying baryons in the CGM of distant galaxies can be done using projected pairs of galaxies, where gas in the halo of the foreground source can leave detectable absorption line features in the spectrum of the background source. The different gas phases of the foreground source can then be studied using absorption lines corresponding to metals in different ionization states. Such studies have been conducted on 8m telescopes using projected QSO-galaxy pairs at $z<2.5$, with only a handful of known host galaxy absorbers at $z>2$. With the ELT, one will be able to conduct the same analysis but now using LBGs as background sources instead of QSOs, which will provide many more lines of sight per field, giving statistics at an epoch which remains out of reach of current facilities.

The IGM tomography programme for MOSAIC (see Sect. 3.3.2) will offer plenty of lines of sight for a spin-off CGM programme without further observations. Preliminary analysis showed that using the sub-sample of LBGs at $z\sim2.8-3.0$ down to $R_{AB}=25$ will offer ~10-15 galaxy-galaxy projected pairs within $\Delta z=0.2$ and a projected distance <200 kpc that could be used to study the CGM of the foreground source. To explore how MOSAIC could allow probes of the warm gas phase in the CGM of such distant galaxies, we sampled the curve of growth for the doublet transition CIV$\lambda\lambda$1548,1550 with a grid of simulations in column density and line broadening ($R=5000$, 20 hr exposure time, see example in Figs. 4 and 5). In the linear regime, we found that MOSAIC can retrieve the gas column density down to a lower limit of $\log[N(CIV)\ cm^{-2}]$ ~13.5 towards background sources as faint as $R_{AB}\sim24.5$. This corresponds to a projected separation <200 kpc in the halo of a typical L* galaxy (Bordoloi 2014). For the saturated line profiles, a lower limit of $\log[N(CIV)\ cm^{-2}]$ ~15.0 can be considered, irrespective of the magnitude of the background galaxy.

Further simulations are on-going to assess which combinations of metallic absorption lines could be recovered in a single observation and spectral set-up, and which gas phases (i.e., cold vs. warm vs. hot) can be studied as a function of redshift. It would be particularly interesting to conduct this programme (as well as the IGM tomography) in the same sample that in which spatially resolved-kinematics will be measured (see Sect. 3.2). Indeed, when combined with the physics extracted from absorption lines in the spectra of background sources, stringent constraints can be obtained on the CGM of these galaxies. In practice, the explored gas phase depends on the covered absorption lines, which is a function of the redshift of the foreground galaxy. The parent sample described in Sect. 3.2 promises statistically large samples of targets in each redshift bin for each gas phase. This would allow estimates of CGM gas profiles in different phases, which will then be used to estimate the CGM mass budget for different gas phases at those redshifts.

---

[2] Using 35% throughput for ELT-MOSAIC in the comparison leads to MOSAIC better than TMT in survey speed by a factor 1.6.

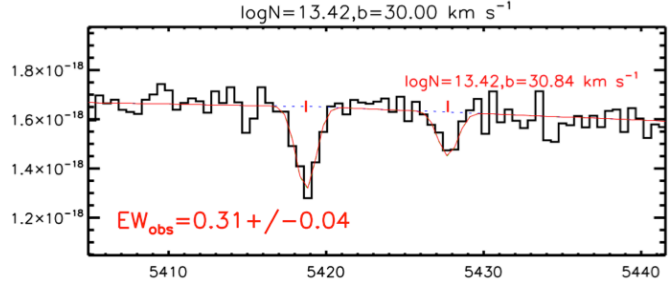

**Fig. 4:** Example of a simulated CIV absorption line in a 23.25 mag background LBG at $z\sim3$. Simulated spectrum extracted from a parent IGM tomography survey using the MOSAIC VIFUs with 20 hr of exposure time per target at $R=5000$. Such a survey would provide a sub-sample of background/foreground galaxy-galaxy pairs without requiring further observational follow-up.

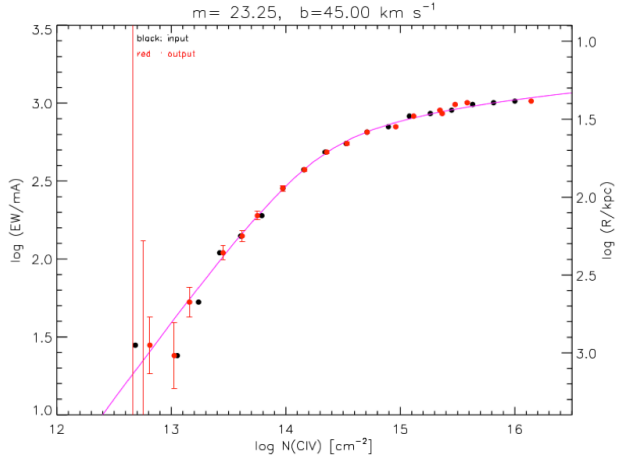

**Fig. 5:** Curve-of-growth of the CIV line reconstructed from simulated observations. Simulated equivalent width (EW) and column densities (N) are shown in black while the measured values from the simulated spectra (similar to that in the left-hand panel but with b=45 km/s instead of 30 km/s) are shown in red. The pink line shows the theoretical relation between the two quantities from which the simulation grid was drawn. The black points lie along this curve but have been offset at the measured EW in the simulation to enable comparison between the input and measured values. The right axis gives the average impact parameter corresponding to a given EW following the correlation found by Bordoloi (2014).

### 3.4. Survey 3: First-light galaxies and reionisation

Identifying the sources of the reionisation of the IGM will be one of the most prominent objectives of *JWST*. Candidate targets will be identified using the drop-out technique from multi-band photometry obtained with *JWST*-NIRCam or even ELT-MICADO. Low-resolution prism spectroscopy from *JWST* will provide an efficient way to measure the redshifts of these candidates and identify the very faint (mag$_{AB}\sim30$ and even fainter), low-luminosity, $z>7$ galaxies that are expected to be responsible for the reionisation process. Spectrographs on the ELT, including MOSAIC, will provide another way to identify such sources and guarantee that the most distant sources will still be observable once the *JWST* mission ends. Simulations (see Fig. 6) show that identifying Lyman-alpha emitters (LAEs) or LBGs at $z\sim7$-9 could be done within $\sim$10 hr exposures (with spectral resolution re-binned down to $R\sim1000$ for the LBGs).

However, understanding the physical processes driving the emission of the UV ionizing photons in these galaxies will require observations at higher spectral resolution than available from *JWST*. In particular, the shape of the Ly$\alpha$ emission line carries important information on the Lyman-continuum leaking process, hence on the escape fraction of Ly$\alpha$ photons in these objects. Coupled to radiative transfer simulations, MOSAIC observations resolving this line at $R=5000$ will provide the required measurements (out of reach of *JWST*-NIRSpec which has $R=2700$ at best). NIRSpec will identify nebular emission lines such as [OII], H$\beta$ and [OIII] out to $z\sim9$. These will be important to calibrate the Ly$\alpha$ photon escape fraction as a function of SFR or metallicity, but NIRSpec will be limited to mag$_{AB}\sim25$ (with $R=1000$, see Fig. 7). Fainter sources will require ELT spectroscopy to measure their UV continua and study the origin of their ionising radiation fields via UV absorption lines and associated diagnostics (see Vanzella et al. 2014 and Fig. 7).

MOSAIC will be the most efficient ELT spectrograph to follow-up *JWST* sources. Integrating the luminosity function between $z=7$ and 9 down to $J_{AB}=30$ (see, e.g., McLure et al. 2013) suggests that ~1600 LBGs should be observable within one MOSAIC field-of-view. Observing all these sources to constrain the evolution of the Lyα luminosity function in three bins of redshifts at $z\sim7$, 8, and 9 would take ~94n with MOSAIC using the 80 HMM-NIR fibre bundles, while a ~60% complete survey (to limit the impact of completeness corrections, see e.g. Drake et al. 2018) would take ~57n. Assuming that the sources already have redshifts from *JWST*, following-up all the identified LAEs to measure the shape of their Lyα emission line at higher spectral resolution ($R$=5000) would be done within a 27n programme, while a 60% complete follow-up would take ~16n. Constructing integrated UV rest-frame spectra using HDM observations of the brightest ($J_{AB}\sim27$) LBGs at $z\sim7$ (20 sources per field) would take an additional ~19n (11n for a 60% complete follow-up). The 80 mas spaxels of the HDM IFUs will provide an optimal surface brightness sensitivity for constructing integrated spectra from such observations (see detailed simulations in Disseau et al. 2014).

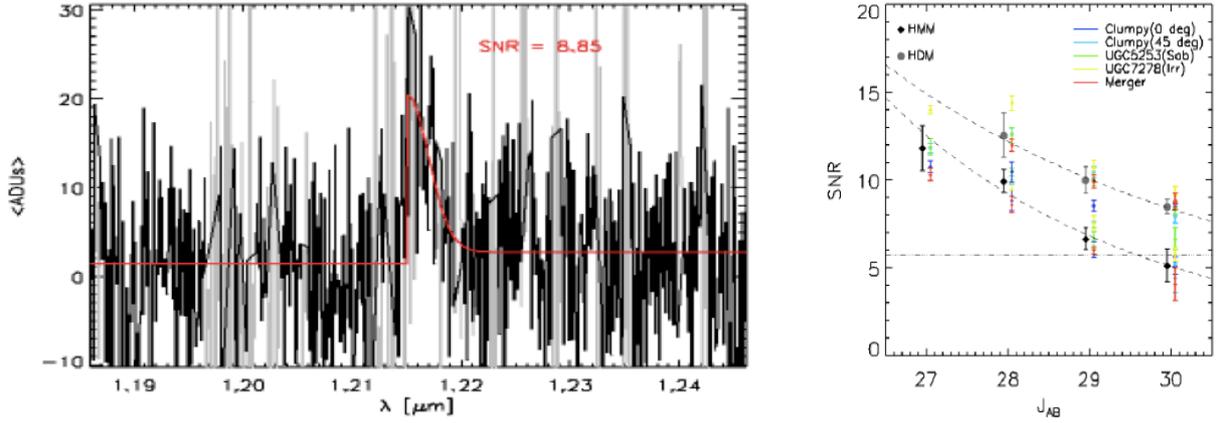

**Fig. 6:** *Left:* simulated integrated spectrum obtained from MOSAIC-HDM observations of a LAE at $z$=9 with $J_{AB}$=28 with 10 hr of integration ($R$=5000). The red line shows the input LAE template. *Right:* expected *S/N* in the Lyα emission line at $z$=9 as a function of magnitude and morphology of the target, for both HDM and HMM-NIR observations. Error bars represent the variation of *S/N* obtained in 33 repeated simulations; horizontal dashed lines represent the requirement in *S/N* per pixel.

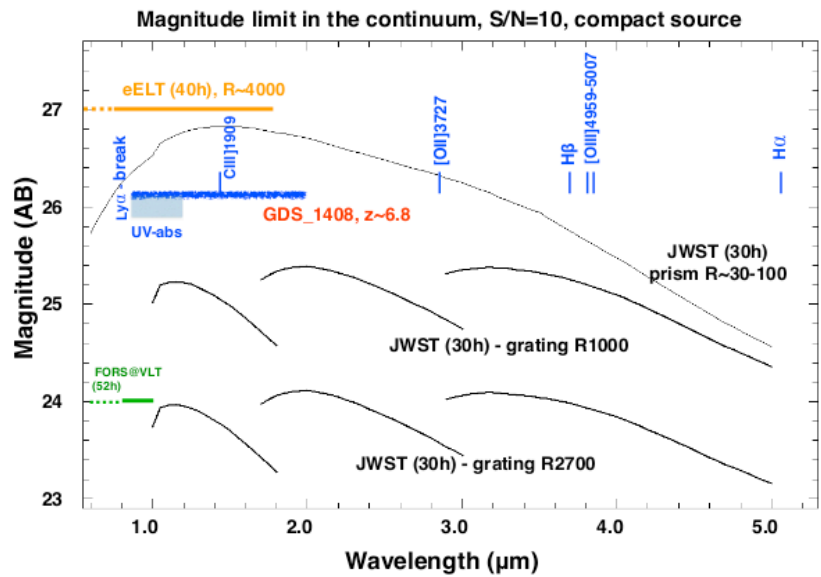

**Fig. 7:** Expected magnitude limits in the continuum with 30–50 hr integrations for ELT-MOSAIC (orange), *JWST* (black) and VLT-FORS (green, from Vanzella et al. 2014 and references therein). The position of the typical UV absorption lines from Lyα to CIV 1550 (grey region, UV-abs) and the basic emission lines, e.g., CIII]1909, [OII]3727, Hβ, [OIII]4959-5007 and Hα are shown.

## 3.5. Survey 4: Extragalactic stellar populations

With its vast primary aperture and excellent angular resolution, the ELT will be the facility to unlock spectroscopy of stellar populations in the broad range of galaxies in the Local Volume, from the edge of the Local Group, out to Mpc distances. There are many compelling target galaxies for stellar spectroscopy of individual resolved stars with the ELT, including the spiral dominated Sculptor 'Group' (at 2-4 Mpc) and the M83/NGC5128 (Cen A) grouping (at ~4-5 Mpc). There is already substantial deep imaging of these galaxies from the *HST* and ground-based telescopes, i.e. we already have catalogues of potential targets, but lack the spectroscopic sensitivity with 8-10m telescopes for sufficient *S/N*. HARMONI will be well suited to spectroscopy of stars in individual regions in external galaxies (and the Milky Way), but the larger samples needed to explore entire galaxy populations will require observations with MOSAIC.

### 3.5.1. Evolved stellar populations

For determinations of stellar metallicities and radial velocities in the old stellar populations beyond 1 Mpc we envisage two different diagnostic approaches. If stellar crowding is not a limiting factor, GLAO-corrected observations will suffice and the primary diagnostic will likely be the calcium triplet (at 0.85-0.87μm); the correlation of its intensity with metallicity is well understood (e.g. Cole et al. 2004; Carrera et al. 2007; Battaglia et al. 2008) and appears robust to metallicities as low as [Fe/H] ≥ −4 (Starkenburg et al. 2010). In denser regions of external galaxies we will require the improved spatial resolution obtained from high-performance AO. In these cases, *J*-band spectroscopy appears an attractive approach for direct metallicity determinations of red stars out to large distances (Davies et al. 2010; Evans et al. 2011; Gazak et al. 2014). Moreover, with AO-corrected IFUs multiple stars can be observed simultaneously per IFU (leading to a large effective multiplex compared to the number of IFUs).

Observations with both modes will be highly complementary, probing different regions and populations (thus providing good sampling of each spatial and kinematic feature). From sensitivity calculations (e.g. Evans et al. 2011; Evans et al. 2015), MOSAIC spectroscopy will provide a huge new number of galaxies for detailed study for the first time, with the capability to determine stellar metallicities out to Mpc distances (for stars near the tip of the RGB) and, in the case of the AO-corrected *J*-band observations, out to tens of Mpc for red supergiants. These will require observing times ~10 hr per pointing, with ~2-10 pointings per galaxy. This programme is particularly flexible since it could be accommodated as a regular programme if one consider only a limited number of pointings and targets, but a more systematic study of the Sculptor group would lead to a more ambitious programme with several tens of nights.

### 3.5.2. Extragalactic massive stars

Spectroscopic surveys in the Milky Way and Magellanic Clouds have provided a wealth of empirical data to improve and refine our understanding of the physics and evolution of massive stars (e.g. Evans et al. 2005). To further our understanding of the role of environment on stellar evolution, particularly at low metallicities, we require high-quality visible spectroscopy (S/N>50, at *R* of a few 1000). Comparisons of such observations with the latest model-atmosphere codes would then yield estimates of physical parameters such as: temperature, luminosity, mass-loss rate, chemical abundances, and rotational velocity. However, in terms of distance, we have reached the limit of what is possible with current large ground-based telescopes – with observations of O-type stars out to ~1.2 Mpc (Tramper et al. 2011, 2014; see also Garcia & Herrero 2013, Camacho et al. 2016). Even then, we are limited to the most luminous (partially evolved) stars. Spectroscopy of main-sequence massive stars beyond the Local Group requires the sensitivity of the ELT and MOSAIC.

To investigate the potential performance of MOSAIC in this field we undertook preliminary simulations of optical spectroscopy of main-sequence O-type stars. This looks promising in terms of high-quality observations of diagnostic lines such as Hα, Hβ and various He I/II lines and will be investigated further this year. Fig. 8 shows the simulation results as a function of *V*-band magnitude. A 10 hr integration delivers S/N>50 down to *V*~22 mag, which would open-up spectroscopy of mid-late O-type dwarfs out to ~3 Mpc, and bright giants/supergiants out to 5-6 Mpc for the first time, encompassing e.g. the large spirals of the Sculptor Group, access to main-sequence massive stars in the dwarfs of the NGC3109 assocation (e.g. Bellazzini et al. 2013), and more.

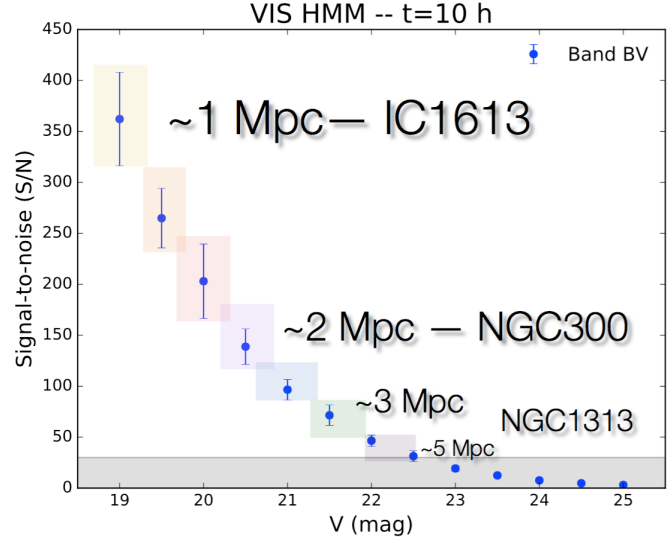

**Fig. 8**: Signal-to-noise (S/N) of simulated O-type spectra vs. *V*-band magnitude; annotations indicate the *S/N* recovered for a mid-late O-type bright giant/supergiant at different distances.

### 3.6. Structure of the extragalactic surveys

#### 3.6.1. Global survey time

The inter-relation of the potential extragalactic surveys is shown in Fig. 9. If conducted in the same fields on the sky, the total requested survey time can be optimised by feeding the different programmes with sub-samples defined from the two survey parent samples. This strategy would lead to a total survey time of ~300 nights to conduct both. Another ~30n would be required to conduct Survey 3 (first-light sources and reionisation), assuming a 60% complete follow-up of *JWST* sources, giving a total of ~330n. Assuming 300n of operations per year split between five instruments and accounting for ~20% of downtime leads to an average of ~50n of observations per year, per instrument. Assuming the instrument baseline described in Sect. 2.3, it would take ~6.6 years of operations for the proposed extragalactic surveys to be completed.

Increasing the MOSAIC multiplex will mostly impact the surveys conducted in the near-IR, since those requiring visible observations are limited by the density of targets and not the multiplex provided by the HMM-Vis mode (see Sect. 3.3). Increasing the multiplex from 80 to 100 in the HMM-NIR and from 8 to 10 IFUs in HDM would result in a gain of 20% in survey speed for the corresponding programmes (see Sect. 3.1) so that the global survey time for extragalactic surveys becomes 100n (Survey 1) + 178n (Survey 2) + 24n (Survey 3) ~ 300n, i.e., a net gain of ~ 10% in terms of survey speed for the extragalactic surveys (requiring 6 years of ELT operations to be completed).

#### 3.6.2. Preliminary considerations on operations

Taking the three extragalactic surveys in Fig. 9 as representative of regular MOSAIC operations provides an interesting illustration of the potential use of MOSAIC in terms of observing modes vs. seeing conditions (see Table 3). Among the four MOSAIC observing modes, HDM and HMM-Vis will represent most of the observing time, almost equally shared. In terms of seeing requirements, HDM observations will require good turbulence conditions corresponding to the first or second quartiles, while HMM-Vis observations have almost no seeing requirements and can be conducted more as a 'filler' programme in the third and fourth quartiles. This means that the three extragalactic surveys do not compete in terms of seeing conditions and can be executed in parallel depending on the turbulence properties. These numbers will have to be revised as the project progresses towards first light to include non-extragalactic surveys and shorter programmes, as defined by the community.

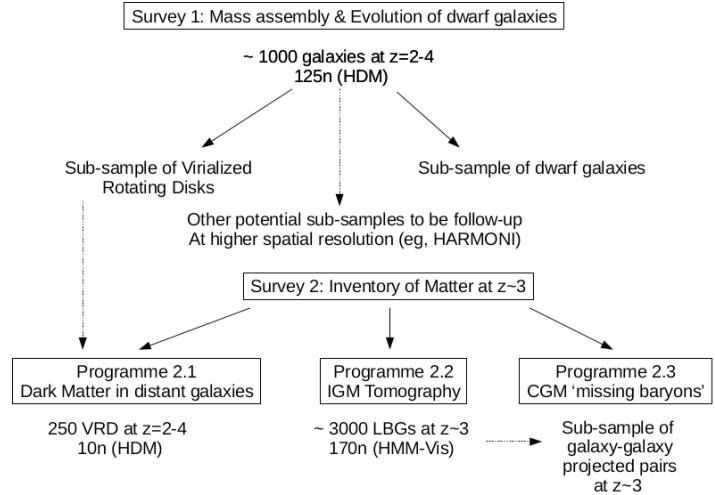

**Fig. 9:** Structure of the first two extragalactic surveys with the estimated survey times.

**Table 3:** Partition of the surveys in terms of observing time and seeing requirements.

| MOSAIC observing mode | Surveys | # of nights | Total # of nights / mode | % of total | Seeing requirement |
|---|---|---|---|---|---|
| HDM | S1:dwarf galaxies<br>S2:inventory of matter<br>S3: first light | 125<br>10<br>11 | 146 | 44% | Q1-Q2 |
| HMM-NIR | S3: first light | 16 | 16 | 5% | Median |
| HMM-Vis | S2: inventory of matter | 170 | 170 | 51% | Q3-Q4 |
|  | TOTAL | 332 |  |  |  |

## 4. ELT-MOSAIC IN THE CONTEXT OF OTHER FUTURE FACILITIES

We now briefly compare the capabilities of MOSAIC with other upcoming facilities to assess the competitiveness of the current concept and the surveys outlined here.

### 4.1. GMT-GMACS

The GMACS concept uses slit masks over a 30-50 arcmin$^2$ field, i.e. a comparable area to that seen by MOSAIC. The notional capabilities are for low- and intermediate-resolution spectroscopy ($R = 1,000\text{-}5,000$) over the range 350-950 nm with a possible extension into the *J*-band (e.g. DePoy et al. 2014). With 50-100 slits/mask, GMACS will provide spectroscopy analogous to the HMM-VIS channels of MOSAIC, but with significantly reduced sensitivity given the collecting area of the GMT is only 37% that of the ELT. We envisage GMACS as a complementary capability once on sky (e.g. spectroscopic confirmation of sources from photometric catalogues, for detailed follow-up at longer wavelengths with MOSAIC). This aspect is even more compelling in light of plans to deploy GMACS behind the MANIFEST fibre positioner, which would feed targets to the spectrograph from an area five to six times larger (using the Starbugs technology developed at the AAO and now in early operation on the UK Schmidt Telescope for the TAIPAN survey, see Lorente et al. these proceedings).

### 4.2. TMT-WFOS

The design of the WFOS instrument has undergone multiple revisions over recent years, but is still focused on visible spectroscopy, from as far blue as possible (opening-up IGM science at lower redshifts) through to ~1 μm, at $R\sim1500$ and 5000 (e.g. Bundy et al., these proceedings). Given the low efficiency of the ELT shortwards of 0.45 μm we have not attempted to compete with TMT on the short-wavelength cases, but at longer wavelengths HMM-VIS observations with MOSAIC will be competitive (see Sect. 3.3.2).

### 4.3. JWST

Beyond considerations such as the fixed mission lifetime of the *JWST,* an important advantage of MOSAIC is its spectral resolving power. The requirement of $R$=5000 is important for the first-light and galaxy evolution cases (Sect. 3.4) enabling studies of the detailed physics in galaxies using strengths and profiles of the ISM lines (e.g. SiIV, OI, CIV) and stellar absorption lines (NV, CIV, SiIV, NIV, with some ISM contribution to these). We highlight that the identification of significant velocity offsets between UV ISM lines, Lyα, and photospheric lines led to the discovery that a large fraction of star-forming galaxies at $z$~3 are driving strong winds ('superwinds'), a feedback process thought to be the dominant mechanism that expels baryons from galaxies at these early times (e.g. Wilman et al. 2005). The likely shallower potential wells of typical galaxies at $z$>7 suggests that outflows from high-$z$ galaxies could have cleared material from the galaxy and surroundings, allowing ionising photons to escape, as well as enriching the IGM with metals. *JWST*-NIRSpec will provide, at best, $R$~2700, strongly limiting study of how the IGM was ionised and enriched.

### 4.4. ELT-HARMONI

HARMONI will provide single IFU observations from the visible through to the *K*-band at a range of spaxel scales, from the diffraction limit (4 mas) to a coarsest scale of 30x60 mas, therefore offering strong synergies with MOSAIC to give access to spatially-resolved measurements at small angular scales. This will enable studies of small-scale structures in distant galaxies (e.g., bars, clumps etc, see Sect. 3.1) as well as probing more crowded regions in studies of stellar populations (see Sect. 3.5). The extension to longer wavelengths than MOSAIC (see Sect. 2.1) will also provide strong complementarity (for single objects) to the galaxy mass assembly case (Sect 3.2). The science cases relying on MOSAIC observations with the near-IR IFUs require large observational samples at moderate spatial resolution (see Sect. 3). In this context, we argue in favour of survey speed rather than pushing the spatial sampling of the MOSAIC HDM IFUs further toward the diffraction limit. This balances the spatial sampling with the surface-brightness detection at angular scales that are essential for each driving science case. The adopted ~80 mas spaxel scale provides an optimal compromise and hence was adopted as a TLR for the Phase A study.

We compare the two instruments in terms of survey speed in the *H*-band in Table 4. As expected, MOSAIC benefits from a net advantage in survey speed ~25 to 50 (depending on $R$), provided mainly by the combination of the multiplex (8 vs. 1), and the larger surface per spaxel (6400 mas$^2$ vs 1800 mas$^2$). Of course, such an advantage stands only for science cases in which the target source density is equal to or larger than the instrument multiplex. In the last line of Table 4, we estimated the survey speed advantage of MOSAIC per IFU, i.e., assuming a multiplex of 1. The larger surface per spaxel still provides a gain of ~ 3-10 per IFU in HDM (again depending on $R$).

Of course, the choice of 80 mas for the HDM spaxel scale renders some science cases more difficult to be conducted with MOSAIC. This is particularly true for cases related to AGN/galaxy coevolution & AGN feedback (SC4), which require spaxel scales closer to ~40 mas to limit beam-smearing effects that can strongly bias estimates of outflow velocities (Huseman et al. 2016; Harrison et al. 2018). While such scales remain accessible at the ELT with HARMONI, the expected spatial density of AGNs down to $H_{AB}$~24 is expected to be ~8 per MOSAIC field (Evans et al. 2015), which means that this science case would significantly benefit from the multiplex of the MOSAIC HDM. However, in the current MOSAIC design (see Jagourel et al. these proceedings), the total number of spaxels imaged on the detectors is conserved regardless of the number of IFUs, spaxel scale, or individual field-of-view (FoV). This means that if the HDM spaxel scale is divided by a factor two it has to be accommodated for in the MOSAIC design by either: (1) keeping the same multiplex by dividing the individual FoV of the HDM IFUs by a factor of four, or (2) keeping the same individual FoV by dividing the number of IFUs by a factor of four, as summarized in Table 5. Option 2b is an alternative in which the resulting 2.5 IFUs from Option 2 are merged into a single IFU with a larger field.

Option 1 maintains the number of IFUs at 8 at the cost of a smaller individual FoV. However, the resulting diameter of the FoV does not provide enough area to allow efficient sky subtraction for high-$z$ galaxies (with typical diameters of ~0.75 arcsec at $z$~3-4, see Evans et al. 2015). Option 2 maintains the individual FoV at the cost of multiplex. However, the resulting multiplex is only 2, i.e., clearly not competitive in terms of objects/on-sky area w.r.t. HARMONI. To further illustrate this, consider Option 2b in which the 2 IFUs are hypothetically merged into a single IFU with a FoV of 2.7 arcsec (diameter). The MOSAIC design is not optimized in this case: the two closest HARMONI spaxel scales are 30x60 mas (6.42x9.12 arcsec FoV), and 20x20 mas (3.04x2.48 arcsec FoV), and MOSAIC would not be competitive in comparison to what HARMONI will offer at first light. Thus, going to smaller spaxel scales for MOSAIC would require a significant revision of the design, which probably implies a switch from fibres to slicers (as studied previously in the EAGLE Phase A study, e.g., Cuby et al. 2010). Another possibility for MOSAIC to provide the ELT with *multi*-IFU observations with ~40 mas/spaxel would be to use dithered observations. Simulations will be required to assess the effective feasibility of this technique with the MOSAIC HDM IFUs, at least on the brightest sources. This would offer another strong synergy between MOSAIC and HARMONI regarding science cases related to AGN/galaxy coevolution and AGN feedback, which is anticipated as an important area of research in the 2020s (e.g., Harrison et al. 2018).

**Table 4:** Survey speed comparison between MOSAIC and HARMONI at 1.65 μm*.

|  | **MOSAIC HDM** | **HARMONI $R$=3500** | **HARMONI $R$=7500** |
|---|---|---|---|
| **Spaxel scale [mas x mas]** | 80x80 | 30x60 | 30x60 |
| $R$ | 5000 | 3500 | 7500 |
| **Multiplex** | 8 (10) | 1 | 1 |
| **Dark current [e/s/pix]** | 0.0041 | 0.0041 | 0.0041 |
| **Read-out noise [e/pix]** | 3 | 3 | 3 |
| **Spectral sampling [pix]** | 5 | 3 | 3 |
| **Spatial sampling along X axis [pix]** | 2 | 4 | 4 |
| **Spatial sampling along Y axis [pix]** | 2 | 2 | 2 |
| **Ensquared Energy per spaxel [%]** | 35 % | 45 % | 45 % |
| *S/N* **ratio (HARMONI / MOSAIC)** |  | 0.6 | 0.4 |
| **Survey speed HARMONI vs. MOSAIC** |  | **0.04 (0.03)** | **0.02 (0.01)** |
| **Survey speed H vs. M** *per IFU* |  | 0.3 | 0.1 |

*All ratios refer to HARMONI/MOSAIC. The resulting noise (sky background + read-out noise + dark current) and object flux ratios are estimated within one spaxel and for an element of spectral resolution, for an integration time DIT=600s (see Evans et al. 2016 for details and assumptions). We applied a linear correction in Ensquared Energy (EE) to the object flux ratio before estimating the resulting *S/N* ratio between MOSAIC and HARMONI to account for the better EE delivered by LTAO (HARMONI) compared to MOAO (MOSAIC). EE=45% per spaxel for the HARMONI LTAO (H band) was assumed (Dr B. Neichel., priv. comm.). The last line gives the survey speed ratio per IFU (i.e., assuming multiplex=1 for MOSAIC). Note that the same comparison in the *J*-band gives approx.. the same results (assuming EE=30% for LTAO vs. EE=24% for MOAO).

**Table 5:** Possible design variations for the MOSAIC HDM.

| **HDM IFU options** | **Spaxel scale size (mas)** | **IFU Diameter (arcsec)** | **# of IFUs** |
|---|---|---|---|
| **Baseline** | 80 | 1.9 | 8 |
| **Option 1** | 40 | 0.95 | 8 |
| **Option 2** | 40 | 1.9 | 2 |
| **Option 2b** | 40 | 2.7 | 1 |

## 4.5. Dedicated MOS instruments on 8-10m class telescopes

As detailed in Sect. 3.3.2, one might consider a dedicated massive MOS on a 8-10m telescope as a potential competitor for MOSAIC. Several such MOS are under development or consideration, including VLT-MOONS (Taylor et al. these proceedings), Subaru-PFS (Tamura et al. these proceedings) and CFHT-MSE (Szeto et al. these proceedings) which are expected to provide multiplexes of 1000, 2400, and ~3500 within patrol fields of ~500, 4500, and 5400 arcmin$^2$, respectively. MOONS is being designed to work primarily in the near-IR, but the two other projects will provide a huge multiplex over a large wavelength range in the visible that extends down to <0.4μm.

One could imagine such a dedicated Visible MOS on the VLT where, in terms of survey speed, the multiplex could, in principle, compensate for the much reduced collecting area compared to the ELT (see scaling relations in Sect. 3.1). The required integration time per target would become very large, with >100 hr for IGM tomography (see Sect. 3.3.2). Recent observations have suggested that integration times as long as 50-80 hr are within the grasp of spectrographs at the VLT (e.g., Vanzella et al. 2014; Pentericci et al. 2018) but it remains to be demonstrated that observations two or three times longer can be achieved without reaching a systematic floor that would limit the *S/N* as one adds more and more exposures to get longer integration times (with *S/N* scaling as $1/\sqrt{N}$ in which N is the number of exposures, as expected in a random-noise limited regime).

Future spectroscopy pushing the VLT even further to its limits will be important to assess the feasibility of this for science cases where the *S/N* requirements are not too demanding. However, in some cases, e.g., absorption-line stellar spectroscopy where *S/N* > 50 is often required, integrating for increasingly longer times on 8-10m telescopes simply will not provide the required observations; we require the unprecedented light-gathering power of the ELT.

### 4.6. Dedicated MOS instruments in the near-IR

Table 6 compares the survey speed of MOSAIC to VLT-MOONS (500 objects, each with dedicated sky fibres) and the potential TMT-IRMS concept (with 46 slits over a ~2x0.6 arcmin field, Mobasher et al. 2010), although the latter is not one of the first-generation instruments planned for the TMT. As in Sect. 3.3.2 and 4.5 for the comparison in the visible, we did not account for slight variations in on-sky apertures or spectral resolution; moreover, as the ELT is a optimized for near-IR observations we did not account for possible variations in global throughput (conversely to Sect. 3.3.2). We also assumed that the instrument multiplex rather than the source density limits the effective multiplex (see Sect. 3.2 and 3.4). Table 6 reveals a clear advantage for MOSAIC, which is driven by the larger collecting surface of the ELT, with MOSAIC being a factor ~3-4 faster than any other ground-based MOS instrument in the near-IR.

**Table 6:** Survey speed comparison between MOSAIC HMM-NIR and other near-IR MOS.

|  | D [m] | Effective Multiplex for LBGs (and patrol fields) | Approx. Survey Speed MOSAIC / other |
|---|---|---|---|
| **MOSAIC** | 39 | 80 (40 arcmin$^2$) | 1.0 |
| **VLT-MOONS** | 8 | 500 (500 arcmin$^2$) | 3.8 (±0.2) |
| **TMT-IRMS** | 30 | 46 (2x0.6 arcmin$^2$) | 2.9 (±0.2) |

### 4.7. The unique parameter space of MOSAIC

We now summarise the characteristics that will make MOSAIC unique in the astronomical landscape of the 2020s.

- **Multi-IFUs in the near-IR:** MOSAIC is the only currently planned facility on a 20-40m telescope that will offer multiple IFUs in the near-IR at *R*=5000 over such a large patrol field. Single-IFU instruments such as *JWST*-NIRSpec will probe similar angular resolution but will be limited to R=2700, while ELT-HARMONI will offer finer spatial scales. Studies of TMT instruments led to the IRMS concept (slits over a smaller field) and the IRMOS concept with 20 IFUs over a ~5 arcmin field (Eikenberry et al. 2006), but these are not part of the first-generation instrument suite (see also Sect. 4.6).

    Only MOSAIC will be able to: (1) detect and resolve the UV spectral features necessary to study the origin of the ionizing photons in all the faintest sources identified by *JWST*; (2) gather a large enough sample of galaxies to cover the whole mass function at *z*=2-4 and offer a sufficiently large sub-sample of virialised rotating disks in which accurate rotation curves can be measured and dark matter profiles derived.

- **High resolution (*R*~20,000) multi-object spectroscopy in the near-IR**: MOSAIC is a unique facility in proposing multiplexed observations at *R*~20,000. This will remain out of reach of *JWST* and other planned facilities. Only MOSAIC will be able to resolve internal motions in distant LMC-like dwarf galaxies at z~2-3.

- **Multi-object spectroscopy at both visible and near-IR wavelengths:** although not simultaneously, MOSAIC will provide multiplexed observations in both the visible (>0.45 μm) and near-IR (<1.8 μm), on the world's largest ELT. The other planned ELT MOS instruments will be limited to either the visible or near-IR. Thus, MOSAIC will be a unique, flexible facility for conducting a complete inventory of matter in galaxies at $z\sim3$, as well as multi-wavelength studies of stellar populations in the local Universe.

We have focused in Phase A on developing large surveys to demonstrate the potentially transformation scientific return of MOSAIC. The above unique instrumental characteristics will also guarantee that shorter, more regular programmes will be conducted on MOSAIC, providing compelling capabilities for the community alongside the surveys described in Sect. 3 (see the individual science cases in the ELT-MOS White Paper). This will make MOSAIC the ideal instrument for optical/near-IR spectroscopic follow-up of faint sources from a diverse range of other facilities, including the *JWST Euclid, WFIRST*, LSST, ALMA, *Athena,* the SKA, and more.

The following gives illustrative examples of short programmes that can be completed within just 10hr of integration:

- **HDM:** measurements of the spatially-resolved properties of Haro11-like dwarf starbursts at $z\sim2$, and of rotation curves of $z\sim4$ galaxies, all with multiplexed observations of 8 (goal: 10) galaxies at the same time.

- **HMM-Vis and HMM-NIR:** measurements of the Lyα forest in ~30-40 LBGs at $z\sim3$ at $R\sim3000$ (HMM-Vis), and redshifts of tens of LAEs or LBGs at $z>7$ (HMM-NIR).

- **HDM/HMM-Vis:** metallicities and kinematics of evolved stars in one MOSAIC pointing in HDM mode for one of the Sculptor galaxies, e.g., sampling a large part of the semi-major axis of NGC 253. With multiple stars per IFU, even one night of observations would yield results for >100 stars to begin exploring abundance gradients etc. Observations with the HMM-Vis (or HMM-NIR depending on the final spectral configuration) could also be used in a comparable amount of time to map the abundances and dynamics of the outer regions via CaT spectroscopy. Similarly with the massive-star populations at Mpc distances (HMM-Vis).

## 5. SUMMARY

There are tremendously exciting scientific opportunities ahead with the promise of MOSAIC on the ELT. During the Phase A study we have performed a detailed requirements analysis of the cases previously assembled for the ELT-MOS White Paper, with these merged to form the TLRs for the conceptual design. Alongside this, we have continued to develop some of the cases that we consider will be highlights of future ELT science, including advanced simulations of MOSAIC observations. From discussions within the Science Team and with the community at the Toledo meeting, there is considerable demand for MOS observations on the ELT, as evidenced by the scale of the possible surveys highlighted in Sect. 3. The four topics listed in Table 1 lend themselves to six sub-programmes (with the inventory of matter entailing IGM tomography, missing baryons, and the high-$z$ rotation curves). A possible approach to the GTO would be to split it over these topics – enabling some of the first breakthroughs in these areas, while also serving as pilot studies/samples for larger surveys. In the longer-term, this could then lead into e.g. three large (~100 night sized) Public Surveys, extending over a number of years of ELT operations after completion of the GTO.

Ahead of Phase B, we will monitor the potential capabilities of the instrument (given the final budget), and then revisit the high-priority cases and their requirements at the start of Phase B. One area that we will look critically at in the coming period is the visible capability. In particular, it was not possible to deliver the required performance for high-resolution observations in the visible – both in terms of resolving power and (the strong wavelength response in) grating efficiency. When combined with the low efficiency in the blue, this significantly weakens some of the stellar cases previously proposed for SC6 (see Sect. 2.2), rendering some of them unfeasible. One area of potential study in Phase B could be to investigate increasing the number of VIFUs, which would boost the efficiency of the IGM and missing baryon cases. Equally, we will look further at inclusion of a HR setting around the CaT in the near-IR spectrographs.

In short, further work on the MOSAIC conceptual design will aim to: (1) further balance the HMM-Vis vs. VIFU modes in terms of multiplex vs. performance and cost; (2) further balance the NIR vs. Vis modes in terms of multiplex with the goal of reaching at least 100 apertures in HMM-NIR and at least 10 IFUs in HDM (see Morris et al. these proceedings); (3) consolidate and finalise the characteristics of the HR spectrograph setups.

Of course, it is certain that compelling new ideas for MOSAIC observations and surveys will arise in the coming years (e.g. from new discoveries with *JWST,* LSST, VLT, and more), so we also need to ensure that we provide discovery space and cover as wide a range of parameter space as we can within the available budget. In this regard, the combination of both the HMM and HDM capabilities – across tens of arcmin$^2$ on the sky – is particularly powerful for future discoveries. It also ensures that MOSAIC will be flexible in terms of scheduling observations with respect to the local conditions and turbulence.